\documentclass[prb,preprintnumbers,amsmath,amssymb,floatfix]{revtex4}

\usepackage{graphicx}
\usepackage{subfigure}
\usepackage{dcolumn}

\makeatletter

\begin{document}
\baselineskip 16pt

\title{Non-relativistic quantum theory at finite temperature}
\author{Xiang-Yao Wu$^{a}$ \footnote{E-mail: wuxy2066@163.com},
Bo-Jun Zhang$^{a}$, Xiao-Jing Liu$^{a}$, Si-Qi Zhang$^{a}$\\ Jing
Wang$^{a}$,Hong Li$^{a}$, Nou Ba$^{a}$, Li Xiao$^{a}$ and Yi-Heng
Wu$^{b}$ }
\affiliation{a.Institute of Physics, Jilin Normal University, Siping 136000 \\
b. Institute of Physics, Jilin University, Changchun 130012}

\begin{abstract}

We propose the non-relativistic finite temperature quantum wave
equations for a single particle and multiple particles. We give
the relation between energy eigenvalues, eigenfunctions,
transition frequency and temperature, and obtain some results: (1)
when the degeneracies of two energy levels are same, the
transition frequency between the two energy levels is unchanged
when the temperature is changed. (2) When the degeneracies of two
energy levels are different, the variance of transition frequency
at two energy levels is direct proportion to temperature
difference.
\\
\vskip 5pt


\end{abstract}
\maketitle

\maketitle {\bf 1. Introduction} \vskip 8pt

Quantum field theory at finite temperature was motivated by the
increasing interest in studying the properties of matter under
extreme conditions as, for example, at very high temperature or
density. The pioneering works joining together the statistical and
quantum field theory were developed mainly by Matsubara [1] in a
non-relativistic context and, the relativistic case by Fradkin
[2]. The finite temperature gauge theories and the problems
concerning to the choice of a physical gauge and its dependence
was analyzed by Bernard [3], in particular, the free
electromagnetic field. Thermodynamics and statistical mechanics
are powerful and vastly general tools. A rather fuller review of
the necessary statistical mechanics may be found in the book by
Fetter and Walecka [4], which also gives a very full account of
non-relativistic finite-temperature field theory. Semiclassical
series were introduced in quantum mechanics by the pioneering
works of Brillouin [5], Kramers [6] and Wentzel [7]. Semiclassical
methods for finite temperature field theories [8-10] also remained
restricted to derivations of the first term of a semiclassical
series [11], even when the problem was reduced to quantum
statistical mechanics [12, 13], viewed as field theory at a point.

In thermodynamic, the thermodynamic quantities such as internal
energy, free energy, volume, pressure, entropy and so on are
effected by temperature, which are described by the first and
second laws of thermodynamics. In quantum statistical, it is only
considered the impact of temperature on the energy level, particle
numbers and probability distribution, and it isn't from the
quantum theory including temperature. In non-relativistic quantum
theory, the Schrodinger equation doesn't include temperature.
Therefore, the full quantum theory and quantum statistical theory
should be included temperature. In this paper, we extend the
Schrodinger equation from zero temperature to finite temperature.
With the finite temperature quantum theory, we can study the
affect of temperature on quantum systems.

\vskip 8pt

{\bf 2. The free energy of thermodynamic system}
\vskip 8pt

For a system constituting of $N$ particles, the free energy is
defined by
\begin{equation}
F=U-TS,
\end{equation}
where $F$, $U$, $S$ and $T$ are the free energy, internal energy,
entropy and temperature of the system, respectively.

In the system, every particle can be in a series of state, we
define $T^{i}_{j}$, $V^{i}_{j}$ are the kinetic energy and
potential energy of the $j-th$ particle in the $i-th$ state,
$V^{i}_{jm}$ is the interaction energy between the $j-th$ and the
$m-th$ particle in the $i-th$ state. The microcosmic internal
energy is

\begin{equation}
U^{i}=\sum^{N_{i}}_{j=1}(T^{i}_{j}+V^{i}_{j})+\sum^{N_{i}}_{j,m}V^{i}_{jm},
\end{equation}
where $N_{i}$ is the particle number of system in the $i-th$
microscopic state, and $V^{i}_{jm}=0 (j=m)$. According to
statistical principle, the macroscopic internal energy of system
is a statistical average value of its microscopic internal energy.
The macroscopic internal energy $U$ is
\begin{eqnarray}
U=\sum^{M}_{i=1}P_{i}U^{i}=\sum^{M}_{i=1}P_{i}(\sum^{N_{i}}_{j=1}(T^{i}_{j}+V^{i}_{j})+\sum^{N_{i}}_{j,m}V^{i}_{jm}),
\end{eqnarray}
where $P_{i}$ is the probability of system in the $i-th$
microscopic state, and $M$ is the microscopic state number of
system.\\
Defining free energy $f^{i}_{j}$, it is the $j-th$ particle in the
$i-th$ state, then the system microscopic free energy in the
$i-th$ state is
\begin{eqnarray}
F^{i}=\sum^{N_{i}}_{j=1}f^{i}_{j},
\end{eqnarray}
the macroscopic free energy of system is a statistical average
value of its microscopic free energy. The macroscopic free energy
$F$ is
\begin{equation}
F=\sum^{M}_{i=1}P_{i}F^{i}=\sum^{M}_{i=1}\sum^{N_{i}}_{j=1}P_{i}f^{i}_{j},
\end{equation}
the microscopic entropy of the $j-th$ particle in the $i-th$ state
$S^{i}_{j}$ is
\begin{equation}
S^{i}_{j}=-k_{B}\ln (\omega^{i}_{j}p^{i}_{j}),
\end{equation}
Where $\omega^{i}_{j}$ is the degeneracy of the $j-th$ particle in
the $i-th$ state, $p^{i}_{j}$ is the probability of the $j-th$
particle in the $i-th$ state and $k_{B}$ is the Boltzmann
constant, and the microscopic entropy of system in the $i-th$
state is
\begin{equation}
S^{i}=\sum^{N_{i}}_{j=1}S^{i}_{j}=-k_{B}\sum^{N_{i}}_{j=1}\ln(\omega^{i}_{j}
p^{i}_{j})
\end{equation}
the macroscopic entropy of system is a statistical average value
of its microscopic entropy. The macroscopic entropy $S$ is
\begin{eqnarray}
S=\sum^{M}_{i=1}P_{i}S^{i}=-k_{B}\sum^{M}_{i=1}\sum^{N_{i}}_{j=1}P_{i}\ln(\omega^{i}_{j}
p^{i}_{j}),
\end{eqnarray}
substituting Eqs. (3), (5) and (8) into (1), we have
\begin{equation}
\sum^{M}_{i=1}\sum^{N_{i}}_{j=1}P^{i}f^{i}_{j}=
\sum^{M}_{i=1}P_{i}(\sum^{N_{i}}_{j=1}(T^{i}_{j}+V^{i}_{j})+
\sum^{N_{i}}_{j,m}V^{i}_{jm})+Tk_{B}\sum^{M}_{i=1}\sum^{N_{i}}_{j=1}P_{i}\ln(\omega^{i}_{j}
p^{i}_{j}),
\end{equation}
Eq. (9) is the macroscopic free energy equation of system.
Deleting the sum mark $\sum^{M}_{i=1}P_{i}$, we have
\begin{equation}
F^{i}=\sum^{N_{i}}_{j=1}f^{i}_{j}=\sum^{N_{i}}_{j=1}(T^{i}_{j}+V^{i}_{j})+\sum^{N_{i}}_{j,m}V^{i}_{jm}+Tk_{B}\sum^{N_{i}}_{j=1}\ln(\omega^{i}_{j}
P^{i}_{j}),
\end{equation}
Eq. (10) is the system's microscopic free energy equation in the
$i-th$ state. Deleting the sum mark $\sum^{N_{i}}_{j=1}$, there is
\begin{equation}
f^{i}_{j}=T^{i}_{j}+V^{i}_{j}+Tk_{B}\ln(\omega^{i}_{j} p^{i}_{j}),
\end{equation}
Eq. (11) is the free energy equation of the $j-th$ single particle
in the $i-th$ microscopic state. Deleting the suffix $j$, we
obtain the free energy equation of arbitrary particle in $i-th$
microscopic state, it is
\begin{equation}
f^{i}=T^{i}+V^{i}+Tk_{B}\ln(\omega^{i} p^{i}).
\end{equation}
\vskip 8pt

{\bf 3. Non-relativistic quantum theory at finite temperature}
\vskip 8pt
In section 2, we give the free energy equation of a
single particle and multiple particles system in the $i-th$ state,
they are shown in Eqs. (10) and (12). Quantizing the Eqs. (10) and
(12), we can obtain the finite temperature quantum wave equation
of single particle and multiple particles. Making the mechanical
quantities in Eq. (12) become the operator:
\begin{eqnarray}
\left \{ \begin{array}{lll} &&
\hat{f^{i}}=i\hbar\frac{\partial}{\partial t}\\&&
\hat{T^{i}}=-\frac{\hbar^{2}}{2m}\nabla^{2}\\ && \hat{V^{i}}=V\\
\end{array}
   \right.,
\end{eqnarray}
we have
\begin{equation}
i\hbar\frac{\partial}{\partial
t}\psi_{i}(\vec{r},t,T)=[-\frac{\hbar^{2}}{2m}\nabla^{2}+V+Tk_{B}\ln(\omega^{i}
p^{i})]\psi_{i}(\vec{r},t,T),
\end{equation}
where $\psi_{i}(\vec{r},t,T)$ is the $i-th$ state wave function.
Eq. (14) is the time-dependent and temperature-dependent quantum
wave equation at finite temperature for a single particle in the
$i-th$ state, which is different from the zero temperature quantum
wave equation, i.e., Schrodinger equation. By the method of
separation variable
\begin{equation}
\psi_{i}(\vec{r},t,T)=\psi_{i}(\vec{r})\phi(T)f(t),
\end{equation}
substituting Eq. (15) into (14), there are
\begin{equation}
i\hbar\frac{df(t)}{dt}=E_{i}(T)f(t),
\end{equation}
\begin{equation}
[-\frac{\hbar^{2}}{2m}\nabla^{2}+V]\psi_{i}(\vec{r})=E_{i}(0)\psi_{i}(\vec{r}),
\end{equation}
\begin{equation}
Tk_{B}\ln(\omega^{i} p^{i})\phi(T)=E'_{i}(T)\phi(T),
\end{equation}
and
\begin{equation}
E_{i}(T)=E_{i}(0)+E'_{i}(T)=E_{i}(0)+Tk_{B}\ln(\omega^{i} p^{i}),
\end{equation}
where $E_{i}(0)$ is the eigenvalue of Schrodinger equation (17) in
the $i-th$ state. The solution of Eq. (16) is
\begin{equation}
f(t)=e^{-\frac{i}{\hbar}E_i(T) t},
\end{equation}
substituting Eq. (20) into (14), there is
\begin{equation}
[-\frac{\hbar^{2}}{2m}\nabla^{2}+V+Tk_{B}\ln(\omega^{i}
p^{i})]\psi_{i}(\vec{r},T)=E_{i}(T)\psi_{i}(\vec{r},T),
\end{equation}
where $\psi_{i}(\vec{r},T)=\psi_{i}(\vec{r})\phi(T)$, $E_{i}(T)$
and $p^{i}$ are corresponding to the eigenfunction, eigenvalues,
and probability in the $i-th$ state. Eq. (21) is the
time-independent and temperature-dependent quantum wave equation.
When the temperature $T=0$, Eq. (21) becomes Schrodinger equation.

The probability $p^{i}$ is
\begin{equation}
p^{i}=\frac{1}{Z(T)}e^{-\beta E_{i}},
\end{equation}
where $Z(T)=\sum_{i}\omega_{i}e^{-\beta E_{i}}$, and
$\beta=\frac{1}{k_BT}$.

For Eq. (19), there are two situation:

(1) When $T=0$
\begin{equation}
E_{i}(T)=E_{i}(0),
\end{equation}

(2) When $T\neq0$, substituting Eq. (22) into (19), we have
\begin{eqnarray}
E_{i}(T)&=&E_{i}(0)+Tk_{B}\ln(\omega^{i} p^{i})\nonumber\\
&=&E_{i}(0)-E_{i}(T)+Tk_{B}\ln\omega^{i}-Tk_{B}\ln Z(T),
\end{eqnarray}
i.e.,
\begin{equation}
E_{i}(T)=\frac{1}{2}E_{i}(0)+\frac{1}{2}Tk_{B}\ln\omega^{i}-Tk_{B}\ln
\sqrt{Z(T)},
\end{equation}

For a dimensional harmonic oscillator, $\omega^{i}=1$, we have
\begin{equation}
E_{i}(T)=\frac{1}{2}E_{i}(0)-Tk_{B}\ln \sqrt{Z(T)},
\end{equation}
where
\begin{equation}
E_{i}(0)=(i+\frac{1}{2})\hbar \omega,
\end{equation}
and
\begin{eqnarray}
Z(T)&=&\sum_i e^{-\beta E_{i}(T)}\nonumber\\
&=&\sum_i e^{-\frac{1}{k_{B}T} (\frac{1}{2}E_{i}(0)-Tk_{B}\ln
\sqrt{Z(T)})}\nonumber\\
&=&\sum_i e^{-\frac{1}{2k_{B}T}E_{i}(0)}.\sqrt{Z(T)},
\end{eqnarray}
to get
\begin{eqnarray}
\sqrt{Z(T)}&=&\sum_i e^{-\frac{1}{2k_{B}T}E_{i}(0)}\nonumber\\
&=&\frac{e^{-\frac{\hbar\omega}{4k_{B}T}}}{1-e^{-\frac{\hbar\omega}{2k_{B}T}}},
\end{eqnarray}
substituting Eqs. (27) and (29)into (26), we obtain
\begin{equation}
E_i(T)=\frac{i+1}{2}\hbar\omega+k_{B}T\ln(1-e^{-\frac{\hbar\omega}{2k_{B}T}}).
\end{equation}
Eq. (30) is the $i-th$ energy level of a dimensional harmonic
oscillator at finite temperature. Its energy level at zero
temperature and finite temperature can be written as
\begin{eqnarray}
E_{n}=\left \{ \begin{array}{lll} && (n+\frac{1}{2})\hbar \omega
 \hspace{1.62in}(T=0)\\&&
\frac{1}{2}(n+1)\hbar\omega+k_{B}T\ln(1-e^{-\frac{\hbar\omega}{2k_{B}T}}) \hspace{0.19in}(T\neq 0)\\
\end{array}
   \right..
\end{eqnarray}

From Eq. (25), we have
\begin{equation}
2(E_{i}(T_{1})-E_{j}(T_{1}))=E_{i}(0)-E_{j}(0)+T_{1}k_{B}\ln\frac{\omega^{i}}{\omega^{j}},
\end{equation}
and
\begin{equation}
2(E_{i}(T_{2})-E_{j}(T_{2}))=E_{i}(0)-E_{j}(0)+T_{2}k_{B}\ln\frac{\omega^{i}}{\omega^{j}},
\end{equation}
Eq. (32) minus (33), there is
\begin{equation}
\nu_{ij}(T_{1}-T_{2})=\nu_{ij}(T_{1})-\nu_{ij}(T_{2})=\frac{k_{B}}{2h}(T_{1}-T_{2})\ln\frac{\omega^{i}}{\omega^{j}},
\end{equation}
where $\nu_{ij}(T)=(E_{i}(T)-E_{j}(T))/h$, which is the transition
frequency from the $i-th$ state to the $j-th$ state.

From Eq. (34), we can obtain the results: (1) when
$\omega^{i}=\omega^{j}$, $\nu_{ij}(T_{1}-T_{2})=0$, i.e., when the
degeneracies of two energy levels are the same, the transition
frequency is unchanged with different temperature. (2) When
$T_{1}\neq T_{2}$ and $\omega^{i}\neq\omega^{j}$,
$\nu_{ij}(T_{1}-T_{2})$ is direct proportion to $T_{1}-T_{2}$.

The time-dependent and temperature-dependent wave function at
$i-th$ state is
\begin{eqnarray}
\psi_{i}(\vec{r},t,T)&=&\psi_{i}(\vec{r}) \phi(T)f(t)\nonumber\\
&=&\psi_{i}(\vec{r}) \phi(T)e^{-i\frac{E_{i}(T)}{\hbar}t}\nonumber\\
&=&\psi_{i}(\vec{r})\phi(T)e^{-\frac{i}{2\hbar}E_{i}(0)t}\cdot
e^{-\frac{i}{2\hbar}Tk_{B}\ln \frac{\omega^{i}}{Z(T)}},
\end{eqnarray}
quantizing Eq. (10), we have
\begin{equation}
i\hbar\frac{\partial}{\partial
t}\psi_{i}(\vec{r}_{1},\vec{r}_{2},\ldots
\vec{r}_{N_{i}},t,T)=[\sum^{N_{i}}_{j=1}(-\frac{\hbar^{2}}{2m_{j}}\nabla_{j}^{2}+
V_{j})+\sum^{N_{i}}_{j,m}V_{jm}+k_{B}T\sum^{N_{i}}_{j=1}\ln(\omega^{i}_{j}
p^{i}_{j})]\psi_{i}(\vec{r}_{1},\vec{r}_{2},\ldots
\vec{r}_{N_{i}},t,T),
\end{equation}
with the identity principle, there is
\begin{equation}
p_{1}^{i}=p_{2}^{i}=\cdots=p_{j}^{i}=\cdots=p_{N_{i}}^{i}=p^{i},
\end{equation}
Eq. (37) becomes
\begin{equation}
i\hbar\frac{\partial}{\partial_{t}}
\psi_{i}(\vec{r}_{1},\vec{r}_{2},\ldots \vec{r}_{N_{i}},t,T)
=[\sum^{N_{i}}_{j=1}(-\frac{\hbar^{2}}{2m_{j}}\nabla_{j}^{2}+V_{j})+
\sum^{N_{i}}_{jm}V_{jm}+k_{B}T\sum^{N_{i}}_{j=1}\ln(
\omega^{i}_{j}p^{i})]\psi_{i}(\vec{r}_{1},\vec{r}_{2},\ldots
\vec{r}_{N_{i}},t,T).
\end{equation}
Eq. (38) is the finite temperature quantum theory of multiple
particles. \vskip 8pt

{\bf 4. Lagrangian function at finite temperature}

\vskip 8pt

For the finite temperature quantum equation (14), its complex
conjugate equation is
\begin{equation}
-i\hbar\frac{\partial}{\partial
t}\psi^{*}_{i}(\vec{r},t,T)=[-\frac{\hbar^{2}}{2m}\nabla^{2}+V+Tk_{B}\ln(
\omega^{i}p^{i})]\psi^{*}_{i}(\vec{r},t,T),
\end{equation}
the Lagrangian function of the finite temperature quantum
equations (14) and (39) can be taken as
\begin{equation}
L_{i}=i\hbar\psi^{*}_{i}\cdot\dot{\psi}_{i}-\frac{\hbar^{2}}
{2m}\nabla\psi^{*}_{i}\cdot\nabla\psi_{i}-V\psi^{*}_{i}\psi_{i}-Tk_{B}\ln(
\omega^{i}p^{i})\psi^{*}_{i}\cdot\psi_{i},
\end{equation}
where $L_{i}$ is the Lagrangian function of the $i-th$ quantum
state.

From Eq. (40), we have
\begin{eqnarray}
 \frac{\partial L_{i}}{\partial
\psi_{i}}=-V\psi^{*}_{i}-Tk_{B}\ln(
\omega^{i}p^{i})\psi^{*}_{i},
\end{eqnarray}
\begin{eqnarray}
 \frac{\partial L_{i}}{\partial
\dot{\psi}_{i}}=i\hbar\psi^{*}_{i},
\end{eqnarray}

\begin{eqnarray}
\frac{\partial L_{i}}{\partial (\frac{\partial \psi_{i}}{\partial
x_{i}})}=-\frac{\hbar^{2}}{2m}\frac{\partial
\psi^{*}_{i}}{\partial x_{i}} ,
\end{eqnarray}

\begin{eqnarray}
\frac{\partial L_{i}}{\partial
\psi^{*}_{i}}=i\hbar\dot{\psi}_{i}-V\psi_{i}-Tk_{B}\ln(
\omega^{i}p^{i})\psi_{i},
\end{eqnarray}
\begin{eqnarray}
 \frac{\partial L_{i}}{\partial
\dot{\psi}^{*}_{i}}=0,
\end{eqnarray}
and
\begin{eqnarray}
\frac{\partial L_{i}}{\partial (\frac{\partial
\psi^{*}_{i}}{\partial
x_{i}})}=-\frac{\hbar^{2}}{2m}\frac{\partial \psi_{i}}{\partial
x_{i}},
\end{eqnarray}
substituting Eqs. (41)-(43) into Lagrangian equation
\begin{equation}
\frac{\partial L_{i}}{\partial
\psi^{*}_{i}}-\frac{\partial}{\partial t}(\frac{\partial
L_{i}}{\partial
\dot{\psi}^{*}_{i}})-\sum^{3}_{i=1}\frac{\partial}{\partial{x_{i}}}(\frac{\partial
L_{i}}{\partial (\frac{\partial \psi^{*}_{i}}{\partial
x_{i}})})=0,
\end{equation}
to get
\begin{equation}
i\hbar\dot{\psi}_{i}-V\psi_{i}-Tk_{B}\ln(
\omega^{i}p^{i})\psi_{i}+\frac{\hbar^{2}}{2m}\sum^{3}_{i=1}\frac{\partial^{2}
\psi_{i}}{\partial x^{2}_{i}}=0.
\end{equation}
Eq. (48) is equation (14)
\begin{equation}
i\hbar\frac{\partial}{\partial{t}}\psi_{i}(\vec{r},t,T)=[-\frac{\hbar^{2}}{2m}\nabla^{2}+V+Tk_{B}\ln(
\omega^{i}p^{i})]\psi_{i}(\vec{r},t,T),
\end{equation}
substituting Eqs. (44)-(46) into Lagrangian equation
\begin{equation}
\frac{\partial L_{i}}{\partial
\psi_{i}}-\frac{\partial}{\partial{t}}(\frac{\partial
L_{i}}{\partial
\dot{\psi}_{i}})-\sum^{3}_{i=1}\frac{\partial}{\partial{x_{i}}}(\frac{\partial
L_{i}}{\partial (\frac{\partial \psi_{i}}{\partial x_{i}})})=0,
\end{equation}
to get
\begin{equation}
-V\psi^{*}_{i}-Tk_{B}\ln(
\omega^{i}p^{i})\psi^{*}_{i}-i\hbar\dot{\psi}^{*}_{i}+\sum^{3}_{i=1}\frac{\hbar^{2}}{2m}\frac{\partial^{2}
\psi^{*}_{i}}{\partial x^{2}_{i}}=0,
\end{equation}
Eq. (51) is equation (39)
\begin{equation}
-i\hbar\frac{\partial}{\partial{t}}\psi^{*}_{i}(\vec{r},t,T)=[-\frac{\hbar^{2}}{2m}\nabla^{2}+V+
Tk_{B}\ln(\omega^{i}p^{i})]\psi^{*}_{i}(\vec{r},t,T).
\end{equation}
When the finite temperature Lagrangian function is taken as the
form of equation (40), we can obtain the finite temperature
quantum wave equations (14) and (39). \vskip 5pt

\newpage

{\bf 5. Conclusion} \vskip 5pt

With the thermodynamic and statistical mechanics, we give the
microscopic free energy of a single particle and multiple
particles system. By quantization, we give the non-relativistic
finite temperature quantum wave equation for a single particle and
multiple particles, and give the relation between energy
eigenvalues, eigenfunctions, transition frequency and temperature.
Otherwise, we give the relation between transition frequency and
temperature and obtain some results: (1) when
$\omega^{i}=\omega^{j}$, $\nu_{ij}(T_{1}-T_{2})=0$, i.e., when the
degeneracies of two energy levels are the same, the transition
frequency is unchanged with different temperature. (2) When
$T_{1}\neq T_{2}$ and $\omega^{i}\neq\omega^{j}$, the variance of
transition frequency $\nu_{ij}(T_{1}-T_{2})$ is direct proportion
to $T_{1}-T_{2}$. The finite temperature quantum theory should be
tested by experiment in the future, and they can be studied
superconductivity mechanism and Bose-Einstein Condensate and so
on.


\begin{thebibliography}{10}


\bibitem{s1}
T. Matsubra, Prog. Theor. Phys. 14, 351 (1955).

\bibitem{s2}
 E. S. Fradkin, Doklady Akad. Nauk. 98 (1954) 47; ibid. 100,
897 (1955).

\bibitem{s3}
 C. W. Bernard, Phys. Rev. D9, 3312 (1974).

\bibitem{s4}
 Fetter A L and Walecka J D 1971 Quantum Theory of Many
Particle Systems (New York: McGraw-Hill).

\bibitem{s5}
 L. Brillouin, Comptes Rendus 183, 24 (1926).

\bibitem{s6}
 H. A. Kramers, Z. Phys. 39, 828 (1926).

\bibitem{s7}
 G. Wentzel, Z. Phys. 38, 518 (1926).

\bibitem{s8}
 C. W. Bernard, Phys. Rev. D 9, 3312 (1974).

\bibitem{s9}
 L. Dolan and R. Jackiw, Phys. Rev. D 9, 3320 (1974).

\bibitem{s10}
 S. Weinberg, Phys. Rev. D 9, 3357 (1974).

\bibitem{s11}
 R. F. Dashen, Shang-keng Ma and R. Rajaraman, Phys. Rev. D
11, 1499 (1975).

\bibitem{s12}
 B. J. Harrington, Phys. Rev. D 18, 2982 (1978).

\bibitem{s13}
 L. Dolan and J. Kiskis, Phys. Rev. D 20, 505 (1979).
\end{thebibliography}
\end{document}